\documentclass[runningheads]{llncs}
\usepackage{graphicx}
\usepackage{amsmath}
\usepackage{amsfonts}
\usepackage{tabularx}
\begin{document}
\newcolumntype{C}[1]{>{\centering\arraybackslash}p{#1}}

\title{Automatic Radiology Report Generation based on Multi-view Image Fusion and \\ Medical Concept Enrichment}
%

\author{Jianbo Yuan\inst{1} \and
Haofu Liao\inst{1} \and
Rui Luo\inst{2} \and
Jiebo Luo\inst{1}}

\authorrunning{J. Yuan et al.}

\institute{Department of Computer Science, University of Rochester \\
\email{\{jyuan10, hliao6, jluo\}@cs.rochester.edu}\\
\and
Futurewei Technologies, Inc., Bellevue, WA 98004, USA\\
\email{rui.luo@futurewei.com}}
\maketitle  

\begin{abstract} 
Generating radiology reports is time-consuming and requires extensive expertise in practice. Therefore, reliable automatic radiology report generation is highly desired to alleviate the workload. Although deep learning techniques have been successfully applied to image classification and image captioning tasks, radiology report generation remains challenging in regards to understanding and linking complicated medical visual contents with accurate natural language descriptions. In addition, the data scales of open-access datasets that contain paired medical images and reports remain very limited. To cope with these practical challenges, we propose a generative encoder-decoder model and focus on chest x-ray images and reports with the following improvements. First, we pretrain the encoder with a large number of chest x-ray images to accurately recognize 14 common radiographic observations, while taking advantage of the multi-view images by enforcing the cross-view consistency. Second, we synthesize multi-view visual features based on a sentence-level attention mechanism in a late fusion fashion. In addition, in order to enrich the decoder with descriptive semantics and enforce the correctness of the deterministic medical-related contents such as mentions of organs or diagnoses, we extract medical concepts based on the radiology reports in the training data and fine-tune the encoder to extract the most frequent medical concepts from the x-ray images. Such concepts are fused with each decoding step by a word-level attention model. The experimental results conducted on the Indiana University Chest X-Ray dataset demonstrate that the proposed model achieves the state-of-the-art performance compared with other baseline approaches.

\end{abstract}

\section{Introduction}
Medical images are widely used in clinical decision-making. For example, chest x-ray images are used for diagnosing pneumonia and pleural effusion. The interpretation of medical images requires extensive expertise and is prone to human errors. Considering the demands of accurately interpreting medical images in large amounts within short times, an automatic medical imaging report generation model can be helpful to alleviate the labor intensity involved in the task. In this work, we aim to propose a novel medical imaging report generation model focusing on radiology. To be more specific, the inputs of the proposed framework are chest x-ray images under different views (frontal and lateral) based on which radiology reports are generated accordingly. Radiology reports contain information summarized by radiologists and are important for further diagnosis and follow-up recommendations.

The problem setting is similar to image captioning, where the objective is to generate descriptions for natural images. Most existing studies apply similar structures including an encoder based on convolutional neural networks (CNN), and a decoder based on recurrent neural networks (RNN)~\cite{vinyals2015show} which captures the temporal information and is widely used in natural language processing (NLP). Attention models have been applied in captioning to connect the visual contents and semantics selectively~\cite{xu2015show}. More recently, studies on radiology report generation have shown promising results. To handle paragraph-level generation, a hierarchical LSTM decoder has been applied to generate medical imaging reports~\cite{XingXJ18} incorporating with visual and tag attentions. Xue \emph{et al.} build an iterative decoder with visual attentions to enforce the coherence between sentences~\cite{XueXLXATH18}. Li \emph{et al.} propose a retrieval model based on extracted disease graphs for medical report generation~\cite{li2019knowledge}. Medical report generation is different from image captioning in that: (1) data in medical and clinical domains is often limited in scales and thus it is difficult to obtain robust models for reasoning; (2) medical reports are paragraphs other than sentences as in image captioning, where conventional RNN decoders such as long short-term memory (LSTM) have issues of gradient vanishing; and (3) generating medical reports requires higher precision when used in practice, especially on medical-related contents, such as disease diagnosis. 

We choose the widely used Indiana University Chest X-ray radiology report dataset (IU-RR)~\cite{FushmanK16} for this task. In most cases, radiology reports contain descriptive findings in the form of paragraphs, and conclusive impressions in one or a few sentences. To address the challenges mentioned above, we aim to improve both the encoder and decoder in the following aspects:

First, we construct a multi-task scheme consisting of chest x-ray image classification and report generation. This strategy has been shown to be successful because the encoder is enforced to learn radiology-related features for decoding~\cite{XingXJ18}. Since the data scale of IU-RR is small, encoder pretraining is important in order to obtain a robust performance. Different from previous studies using ImageNet which is collected for general-purposed object recognition, we pretrain with large scale chest x-ray images from the same domain, namely CheXpert~\cite{irvin2019chexpert}, to better capture domain specific image features for decoding. Second, most of previous studies using chest x-ray images for disease classification and report generation consider the frontal and lateral images from \emph{the same} patient as \emph{two} independent cases~\cite{XingXJ18,wang2017chestx}. We argue that lateral images contain complementary information to frontal images in the process of interpreting medical images. Such multi-view features should be synthesized selectively other than contributing equally (concatenate, mean or sum) to the final results. Moreover, it is likely to generate inconsistent results for the same patient based on images from different views. We propose to synthesize multi-view information by applying a sentence-level attention model, and enforce the encoder to extract consistent features with a cross-view consistency (CVC) loss.

From the decoder side, we use hierarchical LSTM (sentence and word level LSTM) to generate radiology reports. RNN decoders tend to memorize word distributions and patterns which frequently occur in the training data, and thus might produce inaccurate results when the target contents have not been observed, or when multiple patterns share similar distributions given the previous contents. Such limitations significantly hinders the credibility of machine-generated results in practical use, since incorrect medical-related contents can be very misleading. For example, generating ``left-sided pleural effusion'' while the ground truth is ``right-sided pleural effusion''. In addition, the source visual contents stay too far from the targeted word decoder in hierarchical LSTM which makes the generation process more difficult. To address such issues, we explore the semantics conveyed in the textual contents and apply them directly to the word decoder. We first extract frequent medical concepts based on the radiology reports and fine-tune the encoder to recognize such concepts. The obtained medical concepts contain explicit information to accurately generate deterministic medical-related contents, such as diagnosis, locations, and observations.

The main contributions of our work are summarized as follows: (1) to the best of our knowledge, we are the first to employ the latest CheXpert dataset to obtain a more robust encoder for radiology report generation; (2) we selectively incorporate multi-view visual contents with sentence-level attentions and enforce the consistency between different views; (3) we extract and apply medical concepts to the decoder with word-level attentions to enhance the correctness of the medical-related contents; (4) our integrated framework outperforms the state-of-the-art baselines in the experiments, and (5) we visualize uncertain radiographic observations predicted by the encoder to provide an added benefit to direct more expert attention to such uncertainties for further analysis in practice.

\begin{figure}[t]
  \includegraphics[width=\linewidth]{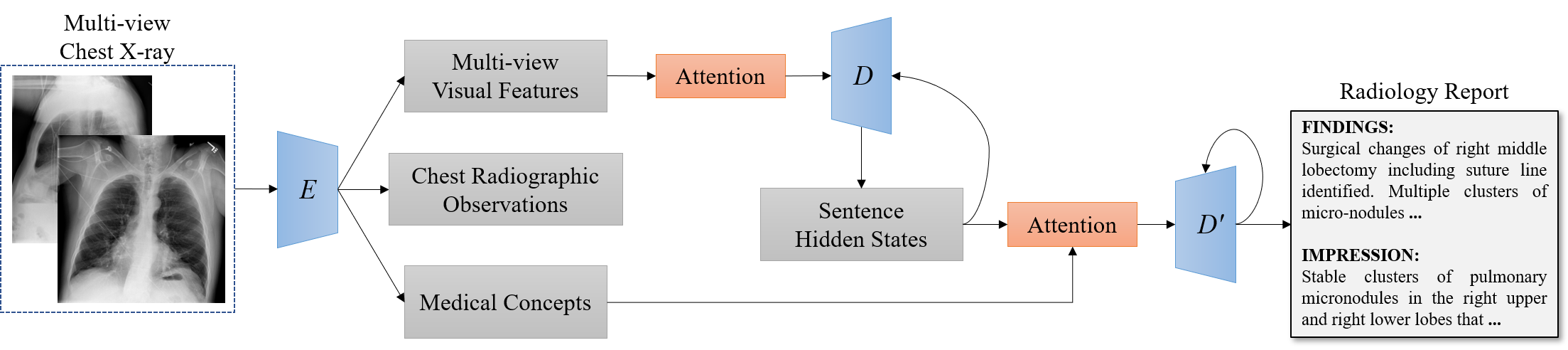}
  \tiny
  \caption{Overall framework of the proposed encoder and decoder with attentions. $E, D$, and $D'$ denote the encoder, sentence decoder, and word decoder, respectively.}
  \label{fig:framework}
\end{figure}

\section{Methodology}
The proposed framework consists of a multi-view CNN encoder and a concept enriched hierarchical LSTM decoder as in Figure \ref{fig:framework}. We apply a multi-task scheme including: (1) radiographic observation classification to pretrain and fine-tune the encoder with large-scale images; (2) to extract medical concepts; and (3) to fuse all information to generate radiology reports. Therefore, two datasets are used in this work: CheXpert~\cite{irvin2019chexpert}, a large collection of chest x-ray images under 14 common chest radiographic observations to pretrain the image encoder, and Indiana University Chest X-ray~\cite{FushmanK16} containing full radiology reports but in a considerably smaller scale for training and evaluating the report generation task.

\subsection{Image Encoder}
The encoder uses Resnet-152~\cite{he2016deep} as the backbone and extracts visual features for predicting chest radiologic observations and radiology report generation. 

\noindent\textbf{Chest Radiographic Observations:}
The task is formulated as a multi-label classification with 14 common radiographic observations following~\cite{irvin2019chexpert} including: \emph{enlarged cardiom, cardiomegaly, lung opacity, lung lesion, edema, consolidation, pneumonia, atelectasis, pneumothorax, pleural effusion, pleural other, fracture, support devices,} and \emph{no finding}. 
Compared with previous studies using pretrained encoders based on ImageNet~\cite{XingXJ18,XueXLXATH18}, pretraining with images from the same domain yields better results. We add one full-connected layer as classifier and compute the binary cross entropy (BCE) loss. Additionally, we consider both frontal and lateral images of one patient as one input pair and enforce the prediction consistencies under different views by a mean square error (MSE) loss over the multi-view encoder outputs. The loss function is thus defined in Equation \ref{eq:1} where $y_{i,j}$ is the $j$-th ground truth entry ($j\in[1,14])$) of the $i$-th sample ($i\in(1, N)$), the frontal view and lateral view prediction are denoted as $\hat{y}_{i_{f}}$ and $\hat{y}_{i_{l}}$. The encoder outputs global features after average-pooling, and local features ${\mathbf{v}\in\mathbb{R}^{k \times d_{v}}}$ from the last CNN block where $k$ denotes the number of local regions and $d_{v}$ denotes the dimension.

\begin{equation}
\label{eq:1}
\mathcal{L}_{I} = -\sum_{v\in \{f,l\}} \sum_{i, j} \left[y_{i,j}\log \hat{y}^{v}_{i,j} + (1-y_{i,j})\log \left(1-\hat{y}^{v}_{i,j} \right) \right] + \lambda \sum_{i} \left( {y}_{i_{f}} - {y}_{i_{l}} \right)^{2}
\end{equation}

\noindent\textbf{Medical Concepts:} The textual reports contain descriptive information related to the visual contents which have not yet been explored by existing models. In IU-RR, Medical text indexer (MTI) can be potentially used in a similar manner~\cite{XingXJ18}. However, MTIs are sometimes noisy and not normalized. Therefore, we use Semrep\footnote{https://semrep.nlm.nih.gov/} to extract \emph{normalized} medical concepts that frequently occur in the training reports. We empirically set the minimal occurrences as 80 and obtained 69 medical concepts for a decent detection accuracy. We fix the pretrained image encoder, and add another fully connected layer on top as the concept classifier.

\subsection{Hierarchical Decoder}

Since conventional RNN decoder is not suitable for paragraph generation, we apply a hierarchical decoder, which has been widely used in paragraph encoding and decoding, to generate radiology reports. The decoder contains two layers: a sentence LSTM decoder that outputs sentence hidden states, and a word LSTM decoder which decodes the sentence hidden states into natural languages. In this way, reports are generated sentence-by-sentence.

\begin{figure}[t]
\centering
  \includegraphics[width=0.92\linewidth]{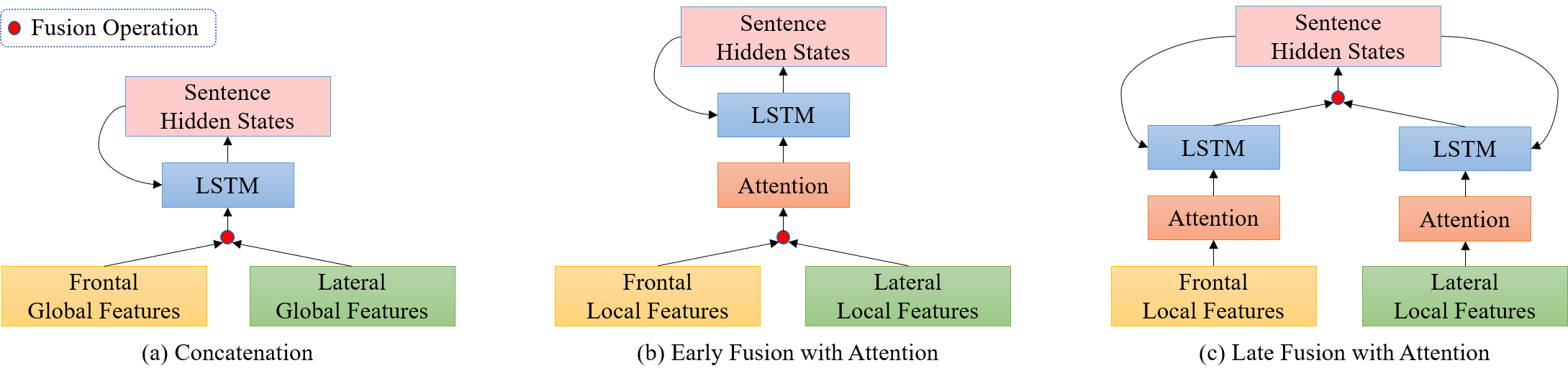}
  \caption{Different fusion schemes for multi-view image features.} 
  \label{fig:fusion}
\end{figure}
\noindent\textbf{Sentence Decoder with Attentions:} The sentence decoder is fed with visual features extracted from the encoder, and generates sentence hidden states. Since we have both frontal and lateral features, the selection of fusion schemes is important. As show in Figure \ref{fig:fusion}, we propose and compare three fusion schemes: an intuitive solution is to directly concatenate the features from both views; early fusion where the concatenated features are selectively attended by the previous hidden state; and late fusion which fuses the hidden states by two decoders after visual-sentence attentions. To generate sentence hidden state $\mathbf{h}_{t_{s}}$ at time step $t_{s}\in (1,N^{s})$, we compute the visual attention weights $\alpha_{i}$ with Equation \ref{eq:att1}, where $v_{m}$ is the $m$-th local feature, and $\mathbf{W}_{a}$, $\mathbf{W}_{v}$ and $\mathbf{W}_{s}$ are weight matrices. 
\begin{equation}
\label{eq:att1}
    \mathbf{a}_{i} = \mathbf{W}_{a}\left[\tanh\left(\mathbf{W}_{v}\mathbf{v_{i}} + \mathbf{W}_{s}\mathbf{h}_{t_{s}-1} \right) \right] \text{,} \quad \alpha_{i}=softmax(\mathbf{a}_{i})
\end{equation}

\noindent By leveraging all the local regions, the attended local feature is thus calculated as $v_{att}=\sum_{m=1}^{k} \alpha_{i,m}v_{m}$, and is concatenated with the previous hidden state to be fed into the sentence LSTM for computing the current hidden state $\mathbf{h}_{t_{s}}$.

\noindent\textbf{Word Decoder with Attentions:} Incorporated with the obtained medical concepts, the sentence hidden states are used as inputs to the word LSTM decoder. 
For each word decoding step, the previous word hidden state $\mathbf{\hat{h}}_{t_{w}}$ for time step $t_{w}\in (1, N^{w}_{t_{s}})$ is used to generate the word distribution over the vocabulary and output the word with the highest score. The embedding $\mathbf{w}_{t_{w}}$ of the predicted word $\hat{w}_{t_{w}}$ is then fused with medical concepts in order to generate the next word hidden state. 
Given the medical concept embeddings $\mathbf{c}\in \mathbb{R}^{n\times d_{c}}$ for $p$ medical concepts for the $i$-th sample, and the predicted concept distributions $\hat{y^{c}_{i}}$, the attention weights over all medical concepts at time step $t_{w}$ is defined in Equation \ref{eq:att2} where $\mathbf{W}_{a^{c}}$, $\mathbf{W}_{c}$, and $\mathbf{W}_{w}$ are the weight matrices to be learned.
\begin{equation}
\label{eq:att2}
    \mathbf{a}^{c}_{i} = \mathbf{W}_{a^{c}}\left[\tanh\left(\hat{y^{c}_{i}}\mathbf{W}_{c}\mathbf{c} + \mathbf{W}_{w}\mathbf{\hat{h}}_{t_{w}-1} \right) \right] \text{,} \quad \alpha^{c}_{i}=softmax(\mathbf{a}^{c}_{i})
\end{equation}

\noindent Similar to visual attention model, the attended medical concept feature is calculated as $c_{att}=\sum_{n=1}^{p} \alpha^{c}_{j,n}c_{n}$, and is concatenated with the previous word embedding to generate the next word. We use cross entropy loss $\mathcal{L}_{W}$ given the predicted word distribution $\hat{y^{w}}_{t_{w}}$ and the ground truth $y^{w}_{t_{w}}$ using Equation \ref{eq:loss2}.
\begin{equation}
\label{eq:loss2}
    \mathcal{L}_{W} = -\sum^{N^{s}}_{t_{s}=1} \sum^{N^{w}_{t_{s}}}_{t_{w}=1} y^{w}_{t_{w}}\log \left(\hat{y^{w}}_{t_{w}} \right)
\end{equation}

\section{Experiment}
\label{exp}
\noindent\textbf{Data Collection:}
\label{data} 
CheXpert~\cite{irvin2019chexpert} contains 224,316 multi-view chest x-ray images from 65,240 patients of 14 common radiographic observations. The observations are generated using NLP tools from the radiology reports labeled as positive, negative, and uncertain. We inherited and visualized the uncertain predictions to address more expert attention for practical use. An alternative dataset is ChestX-ray14~\cite{wang2017chestx}. We chose to use CheXpert because its labeler is reported to be more reliable as compared with ChestX-ray14~\cite{irvin2019chexpert}. 

Since neither of the aforementioned datasets released radiology reports, we use IU-RR~\cite{FushmanK16} for evaluating radiology report generation. For preprocessing, we first removed samples without multi-view images, and concatenated the ``findings'' and ``impression'' sections because in some forms all contents are either in the ``findings'' or ``impression'' section with the other left blank. We filtered out the reports with less than 3 sentences. In the end, we obtained 3,074 samples with multi-view images of which 20\% (615 samples/1,330 images) are used for testing, and the 80\% (2459 samples/4,918 images) are used for training and validation. For encoder fine-tuning, we extract the same 14 labels as~\cite{irvin2019chexpert} on IU-RR. For report parsing, we converted the texts to tokens, and added ``$\langle$start$\rangle$'' and ``$\langle$end$\rangle$'' to the beginning and end of each sentence, respectively. Low frequency (less than 3 occurrences) words were dropped, and textual errors were replaced with ``$\langle$unk$\rangle$'' which are caused by being falsely recognized as confidential information during the original data de-identification of IU-RR.

\begin{table}
\centering
\footnotesize
\caption{Average ROC-AUC (avg-AUC) on radiographic observation classification.}
    \begin{tabular}{C{1.6cm}|C{1.6cm}C{1.6cm}C{1.6cm}C{1.6cm}C{2.1cm}}
    \hline
     Metrics & Base & ImgNet & CX & CX+CVC & CX+CVC+F \\ \hline
    avg-AUC & 0.727 & 0.731 & 0.747 & 0.751 & \textbf{0.764} \\ \hline
    \end{tabular} \label{tab:class}
\end{table}

\noindent\textbf{Chest Radiographic Observations:}
\label{exp:tag}
We conducted extensive experiments on the encoder regarding two factors: how to properly pretrain and fine-tune the encoder, and how to leverage the multi-view information. The classification results on radiographic observations are shown in Table \ref{tab:class}. In general, pretraining with ImageNet (ImgNet) performs marginally better than models without pretraining (Base), and encoders pretrained by CheXpert (CX) performs the best, indicating that pretraining with large scale data in the same domain helps. Enforcing cross-view consistency (CX+CVC) also improves the results. We obtained the best result by fusing multi-view predictions with max operation (CX+CVC+F).

\begin{table}[t]
\caption{Evaluations of generated radiology reports.}
\centering
\footnotesize
\begin{tabular}{C{3.8cm}|cccccc} 
\hline
Methods  & BLEU-1 & BLEU-2 & BLEU-3 & BLEU-4 & METEOR & ROUGE \\ \hline
Vis-Att~\cite{xu2015show}  & 0.399  & 0.251  & 0.168  & 0.118  & 0.167  & 0.323 \\
MM-Att~\cite{XueXLXATH18}  & 0.464  & 0.358  & 0.270  & 0.195  & 0.274  & 0.366 \\
KERP~\cite{li2019knowledge} & 0.482 & 0.325 & 0.226 & 0.162 & $-$ & 0.339 \\
Co-Att~\cite{XingXJ18}   & 0.517  & \textbf{0.386}  & 0.306  & 0.247  & 0.217  & 0.447
\\ \hline
(Ours) MvH   & 0.478 & 0.334 & 0.277 & 0.191 & 0.265 & 0.318 \\
(Ours) MvH+AttE & 0.483 & 0.337 & 0.285 & 0.228 & 0.282 & 0.335 \\
(Ours) MvH+AttL  &  0.488 &  0.357  &  0.296 &  0.246  &  0.313 & 0.351 \\
(Ours) MvH+AttL+MC & \textbf{0.529} &  0.372 &  \textbf{0.315} &  \textbf{0.255} &  \textbf{0.343} &  \textbf{0.453} \\ \hline
(Ours) MvH+AttL+MC* & 0.649 &  0.500 &  0.413 & 0.303 &  0.418 & 0.496 \\ \hline

\end{tabular} \label{tab:cap}
\end{table}

\noindent\textbf{Radiology Report Generation:}
\label{exp:cap}
The evaluation metrics we use are BLEU~\cite{papineni2002bleu}, METEOR~\cite{denkowski2014meteor}, and ROUGE~\cite{lin2004rouge} scores, all of which are widely used in image captioning and machine translation tasks. We compared the proposed model with several state-of-the-art baselines: (1) a visual attention based image captioning model  (Vis-Att)~\cite{xu2015show}; 
(2) radiology report generation models, including a hierarchical decoder with co-attention (Co-Att)~\cite{XingXJ18}, multimodal generative model with visual attention (MM-Att)~\cite{XueXLXATH18}, and knowledge-drive retrieval based report generation (KERP)~\cite{li2019knowledge}; and (3) the proposed multi-view encoder with hierarchical decoder (MvH) model, the base model with visual attentions and early fusion (MvH+AttE), MvH with late fusion fashion (MvH+AttL), and the combination of late fusion with medical concepts (MvH+AttL+MC). MvH+AttL+MC* is an oracle run based on \emph{ground-truth} medical concepts and considered as the upper bound of the improvement caused by applying medical concepts. As shown in Table \ref{tab:cap}, our proposed models generally outperform the state-of-the-art baselines. Compared with MvH, multi-view feature fusions by attentions (AttE and AttL) yield better results. Applying medical concepts significantly improve the performance especially on Meteors because the recalls rise with more semantical information provided directly to the word decoder, and Meteor weights more on recalls over precisions. However, the improvement is subject to prediction errors on medical concepts, indicating that a better encoder would benefit the whole model by a large margin as shown in MvH+AttL+MC*.

\begin{figure}[t]
\centering
  \includegraphics[width=0.9\linewidth]{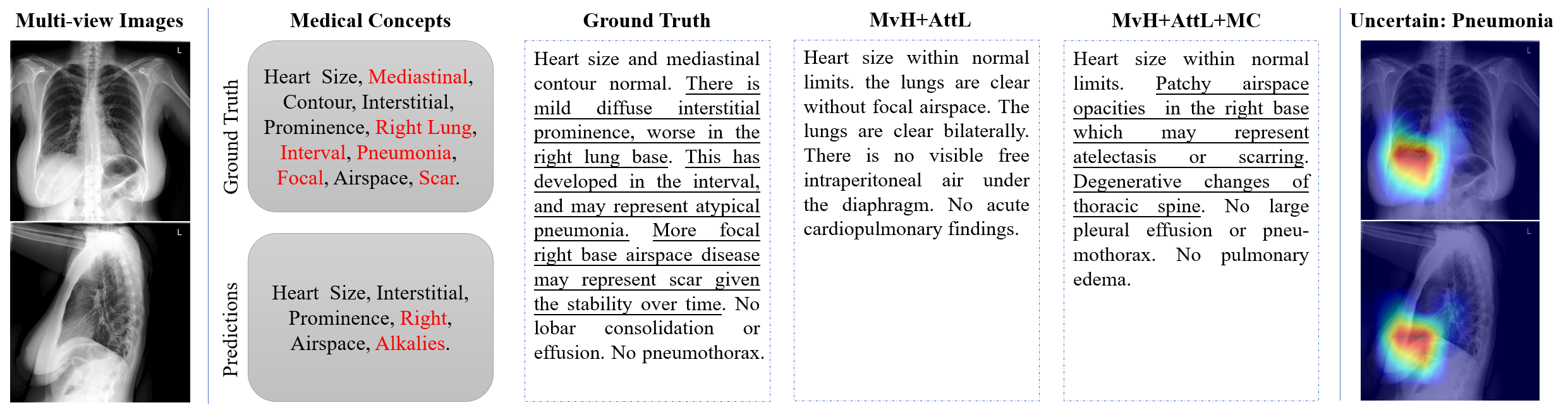}
  \caption{An example report generated by the proposed model. The medical concepts marked red are false (positive/negative) predictions. The underlined sentences are abnormality descriptions. Uncertain predictions are visualized using Grad-cam~\cite{selvaraju2017grad}.}
  \label{fig:example}
\end{figure}

\noindent\textbf{Discussion:} As Figure \ref{fig:example} shows, AttL (and other baseline models) have difficulties generating abnormalities and locations because there is no explicit abnormal information involved in word-level decoding compared with our proposed model. Not all the predicted medical concepts would necessarily appear in the generated reports. On the other hand, the prediction errors from the encoder propagate, such as predicting ``right'' instead of ``right lung'', and affect the generated reports, suggesting a more accurate encoder is beneficial. Moreover, since there are no constraints on the sentence decoder during the training, it is likely to generate similar hidden states for our model. In this case, a stacked attention mechanism would be beneficial for forcing the decoder to focus on different image sub-regions. In addition, we observe that it is difficult for our model to generate unseen sentences and sometimes there are syntax errors. Such errors are due to the limited corpus scale of IU-RR, and we expect by exploring unpaired textual data for pretraining the decoder would address such limitations~\cite{feng2019unsupervised}.

\section{Conclusions}
In this paper, we present a novel encoder-decoder model for radiology report generation. The proposed model takes advantage of multi-view information in radiology by applying visual attentions in a late fusion fashion, and enriches the semantics involved in the hierarchical LSTM decoder with medical concepts. Consequently, both the visual and textual contents have been better exploited to achieve the state-of-the-art performance. The automatic interpretation approach will simplify and expedite the conventional process of generating radiology reports and better assist human experts in decision making. As a valuable added benefit, uncertain radiographic observations are extracted and visualized by our model because it is important to direct more expert attention to such uncertainties for further analysis in practice.

\subsubsection{Acknowledgment}
This work is supported in part by NSF through award IIS-1722847, NIH through the Morris K. Udall Center of Excellence in Parkinson's Disease Research, and our corporate sponsor.

\bibliographystyle{splncs04}
\bibliography{refs}

\begin{thebibliography}{10}
\providecommand{\url}[1]{\texttt{#1}}
\providecommand{\urlprefix}{URL }
\providecommand{\doi}[1]{https://doi.org/#1}

\bibitem{FushmanK16}
Demner{-}Fushman, D., Kohli, M.D., Rosenman, M.B., Shooshan, S.E., Rodriguez,
  L., Antani, S.K., Thoma, G.R., McDonald, C.J.: Preparing a collection of
  radiology examinations for distribution and retrieval. {JAMIA}
  \textbf{23}(2),  304--310 (2016)

\bibitem{denkowski2014meteor}
Denkowski, M., Lavie, A.: Meteor universal: Language specific translation
  evaluation for any target language. In: Proceedings of the ninth workshop on
  statistical machine translation. pp. 376--380 (2014)

\bibitem{feng2019unsupervised}
Feng, Y., Ma, L., Liu, W., Luo, J.: Unsupervised image captioning. In:
  Proceedings of the IEEE Conference on Computer Vision and Pattern
  Recognition. pp. 4125--4134 (2019)

\bibitem{he2016deep}
He, K., Zhang, X., Ren, S., Sun, J.: Deep residual learning for image
  recognition. In: Proceedings of the IEEE conference on Computer Vision and
  Pattern Recognition. pp. 770--778 (2016)

\bibitem{irvin2019chexpert}
Irvin, J., Rajpurkar, P., Ko, M., Yu, Y., Ciurea-Ilcus, S., Chute, C.,
  Marklund, H., Haghgoo, B., Ball, R., Shpanskaya, K., et~al.: Chexpert: A
  large chest radiograph dataset with uncertainty labels and expert comparison.
  arXiv:1901.07031  (2019)

\bibitem{XingXJ18}
Jing, B., Xie, P., Xing, E.P.: On the automatic generation of medical imaging
  reports. In: Proceedings of the 56th Annual Meeting of the Association for
  Computational Linguistics, {ACL} 2018, Melbourne, Australia. pp. 2577--2586
  (2018)

\bibitem{li2019knowledge}
Li, C.Y., Liang, X., Hu, Z., Xing, E.P.: Knowledge-driven encode, retrieve,
  paraphrase for medical image report generation. arxiv:1903.10122  (2019)

\bibitem{lin2004rouge}
Lin, C.Y.: Rouge: A package for automatic evaluation of summaries. In:
  Proceedings of the ACL-04 Workshop. pp. 74--81. Association for Computational
  Linguistics, Barcelona, Spain (July 2004)

\bibitem{papineni2002bleu}
Papineni, K., Roukos, S., Ward, T., Zhu, W.J.: Bleu: a method for automatic
  evaluation of machine translation. In: Proceedings of the 40th annual meeting
  on association for computational linguistics. pp. 311--318 (2002)

\bibitem{selvaraju2017grad}
Selvaraju, R.R., Cogswell, M., Das, A., Vedantam, R., Parikh, D., Batra, D.:
  Grad-cam: Visual explanations from deep networks via gradient-based
  localization. In: Proceedings of the IEEE International Conference on
  Computer Vision. pp. 618--626 (2017)

\bibitem{vinyals2015show}
Vinyals, O., Toshev, A., Bengio, S., Erhan, D.: Show and tell: A neural image
  caption generator. arxiv:1411.4555  (2015)

\bibitem{wang2017chestx}
Wang, X., Peng, Y., Lu, L., Lu, Z., Bagheri, M., Summers, R.M.: Chestx-ray8:
  Hospital-scale chest x-ray database and benchmarks on weakly-supervised
  classification and localization of common thorax diseases. In: Proceedings of
  the IEEE Conference on Computer Vision and Pattern Recognition. pp.
  2097--2106 (2017)

\bibitem{xu2015show}
Xu, K., Ba, J., Kiros, R., Cho, K., Courville, A., Salakhudinov, R., Zemel, R.,
  Bengio, Y.: Show, attend and tell: Neural image caption generation with
  visual attention. In: arxiv:1502.03044 (2015)

\bibitem{XueXLXATH18}
Xue, Y., Xu, T., Long, L.R., Xue, Z., Antani, S.K., Thoma, G.R., Huang, X.:
  Multimodal recurrent model with attention for automated radiology report
  generation. In: Medical Image Computing and Computer Assisted Intervention
  2018, Granada, Spain, Proceedings, Part {I}. pp. 457--466 (2018)

\end{thebibliography}
\end{document}